\documentclass[showpacs, twocolumn]{revtex4-1}

\usepackage{graphicx}
\usepackage{amsmath}
\usepackage{appendix}
\usepackage{bbold}
\usepackage{csquotes}

\begin{document}

\title{Nonequilibrium quench dynamics of Bose-Einstein condensates of microwave-shielded polar molecules}

\author{Abdel\^{a}ali Boudjem\^{a}a}

\affiliation{Department of Physics, Faculty of Exact Sciences and Informatics, and Laboratory of Mechanics and Energy, 
Hassiba Benbouali University of Chlef, P.O. Box 78, 02000, Ouled-Fares, Chlef, Algeria.}
\email {a.boudjemaa@univ-chlef.dz}
\date{\today}

\begin{abstract}
We theoretically investigate the non-equilibrium dynamics of homogeneous ultracold Bose gases of microwave-shielded polar molecules 
following a sudden quench of the scattering length at zero temperature.
We calculate in particular the  quantum depletion, the anomalous density, the condensate fluctuations, and  the pair correlation function using both 
the time-dependent Bogoliubov approach and the self-consistent time-dependent Hartree-Fock-Bogoliubov approximation.
During their time evolution, these quantities exhibit slow or fast oscillations depending on the strength of the shielding interactions.
We find that  at long time scales the molecular condensate is characterized by nonequilibrium steady-state momentum distribution functions, 
with depletion, anomalous density and correlations that deviate from their corresponding equilibrium values.
We demonstrate that the pair correlations expand diffusively at short times while they spread ballistically at long times.
\end{abstract}


\maketitle

\section{Introduction}

Ultracold gases of polar molecules offer an ideal platform for studying many aspects of quantum many-body physics including
quantum computing \cite{DeMille,Wei, Hug,KK,Asn,Saw}, quantum simulation \cite{Blac, Corn}, quantum sensing \cite{DeMille1, Boudj1}, 
quantum magnetism \cite{Gors,Haz}, and precision measurement \cite{Flam, Isaev, Hud} due to their rotational states and controllable long-range anisotropic dipole-dipole interactions (DDIs). 

These achievements have opened longed-for experimental access to the creation of degenerate Fermi gases \cite{Schind} and Bose-Einstein condensates (BECs) \cite{Big}
of microwave-shielded polar molecules (MSPMs).  
Microwave shielding also allows to achieve tetratomic (NaK)$_2$ molecules \cite{Chen}.
Most recently, the ground-state properties of these long-lived ultracold molecules  have been investigated in \cite{WJ} using  a variational wavefunction with Jastrow correlations.
The interaction potential between two MSPMs is characterized by a long-range DDI and a $1/r^6$-type short-range shielding core featuring 
both long-range attractive and repulsive interactions, paving the way to study exotic quantum states such as: 
$p$-wave superfluidity of a microwave-shielded molecular gas \cite{Deng} and molecular droplets stabilized by the two-body shielding potential \cite{WJ, Lang}.
The result of these studies revealed that the Gross-Pitaevskii equation is invalid to treat ultracold molecular condensates \cite{WJ, Lang}.
However,  many questions remain open and unsolved in this field even from a theoretical point of view. 
The major issue concerns the nonequilibrium dynamics of such ultracold gases of bosonic MSPMs.
The long lifetimes of these exotic composites \cite{Big, Chen,Deng} enables us to explore the non-equilibrium dynamics and coherence of the system.

The aim of this work is then to study the dynamics of homogeneous weakly-correlated molecular BECs following a sudden quench of the scattering length at zero temperature.
This complex dynamics can be probed by absorption imaging. 
A stable weakly-correlated  BEC of MSPMs can be reached due to high tunability of the intermolecular potential \cite{Chen, Deng,WJ}.
We focus on the evolution of the quantum fluctuations and the correlations at both short and long times.
Within the time-dependent Bogoliubov approach (TDBA),  we calculate the condensate depletion, the anomalous density, the quantum fluctuations and the pair correlation functions
and look at their spatio-temporal evolution after a sudden switching on of the MSPM interactions.
The TDBA has been extensively employed to study quench dynamics of both dipolar and nondipolar atomic BECs (see e.g. \cite{Natu, Yin, Kain, Natu1, Pav, Cormack}).

In the equilibrium case we show that the condensate depletion and the anomalous density are significantly increased as the size of the shielding core of the two-body potential increases,
leading to a reduction of the condensate fraction. 
We also find that the equation-of-state (EoS) changes its nature from repulsive to attractive depending on the shielding strength unveiling a
phase transition between a gas and a self-bound liquid. 
As the shielding term increases, the correlations develop strong oscillations destroying the long-range order eventually.

However, when the interaction is quenched, the initial Bogoliubov dispersion relation differs from that of the final state. 
Therefore, the system is driven out of equilibrium, such that the normal and anomalous momentum distributions and pair correlations evolve with time.
To deeply understand the role of interactions, we analyze two types of quenches namely: (i) a sudden quench from a noninteracting system
to an interacting molecular BEC and (ii) an instantaneous change from a certain initial microwave-dressed molecular potential to a final value.
Our numerical simulations reveal that the quantum depletion, the anomalous density, the condensate fluctuation, and the second-order correlation function
exhibit slow and fast oscillations depending on the strength of the shielding interaction. 
The spreading of correlations after a quantum quench is also discussed.
We find that correlations display a diffusive motion at short-range distance while they propagate ballistically at long-range distance in the long time regime.

Finally,  we compare our findings with those of the time-dependent Hartree-Fock-Bogoliubov (TDHFB) approach, 
where the condensate is coupled selfconsistently with the noncondensed density and the anomalous density  \cite{Boudj5,Boudj6, Yuk, Boudj7}.  
It is found that the inclusion of the higher-order quantum fluctuations via the HFB theory may alter the quench dynamics of the system.

The rest of the paper is structured as follows.
In Sec.~\ref{model}, we introduce the Hamiltonian of the system and the properties of the molecular interaction potential.
Section \ref{Bogo} deals with the equilibrium properties of ultracold Bose gases of MSPMs. 
The depletion, the anomalous density, the EoS, and the pair correlation function are calculated analytically and numerically using the static Bogoliubov calculation.
In Sec.~\ref{QDyn}, within the TDBA we study the dynamics of uniform weakly-correlated molecular BECs by considering two experimentally feasible quenches.
Analytical expressions governing the dynamics of the noncondensed and anomalous densities, quantum fluctuations and the correlations are derived in detail.
The asymptotic solution of the molecular BEC at long times is also examined.
Section \ref{HFB} discusses the nonequilibrium properties of molecular BECs employing the self-consistent TDHFB approach.
The obtained findings are compared with those of the TDBA.
In Sec.~\ref{conc}, we present our conclusions and discuss possible future developments.

\section{Model}  \label {model}

We consider a uniform three-dimensional (3D) gas of dipolar bosonic molecules at zero temperature which can be produced in a box-shaped trap.
The Hamiltonian of the system is given by
\begin{align}\label{ham}
&\hat H = \int d^3r \, \hat \psi^\dagger ({\bf r}) \left(\frac{-\hbar^2 }{2m}\nabla^2\right)\hat\psi(\mathbf{r}) \nonumber \\
&+\frac{1}{2}\int d^3r\int d^3r^\prime\, \hat\psi^\dagger(\mathbf{r}) \hat\psi^\dagger (\mathbf{r^\prime}) V(\mathbf{r}-\mathbf{r^\prime})\hat\psi(\mathbf{r^\prime}) \hat\psi(\mathbf{r}) ,
\end{align}
where $\hat \psi^\dagger$ and $\hat\psi$ denote the usual creation and annihilation field operators, respectively, and the interaction potential 
which is a combination of an anisotropic Van der Waals-like shielding core and a modified dipolar interaction, reads \cite{Deng,WJ}
\begin{equation}\label{dd}
V(\mathbf r) =\frac{C_3}{r^3} \left(3\cos^2\theta-1\right)+\frac{C_6}{r^6} \sin^2\theta\left(\cos^2\theta+1\right),
\end{equation}
where $\theta$ is the polar angle of $\mathbf r$, and $C_3=  d^2/[48\pi \epsilon_0 (1+\delta_r^2)]$, $C_6=  d^4/\left[128\pi^2 \epsilon_0^2 \Omega \left(1+\delta_r^{3/2}\right)\right]$
are interaction strengths tunable via the Rabi frequency, $\Omega$, and detuning of the microwave field, $\delta$, with $\delta_r= |\delta|/\Omega$ being the relative detuning. 
Here $d$ is the permanent dipole moment in the molecular frame and $\epsilon_0$ is the vacuum permittivity.
The $C_6$ term represents the shielding potential. 
The role of such a microwave shielding term is to provide a repulsive interaction that prevents molecules against short-range losses.
For a circularly polarized microwave, the resulting modified DDI along the $z$-axis represented by the $C_3$ term takes the form 
$V(\mathbf r) =C_3\left(3\cos^2\theta-1\right)/ r^3$ \cite{Deng, Lang} with an extra negative sign. 
This modification may alter the ground-state and the dynamic properties of the system. 

Additionally, there is a short-range interaction potential which is characterized by the $s$-wave scattering length, $a$.
To calculate such a quantity and relate it to the parameters of the potential at hand, one should analyze the long-range behavior of the $s$-wave component 
of the wavefunction at low energy \cite{Yi, Ronen,Lang}. This is indeed not an easy task due to the contribution of the modes $l$, $l\pm2$, and $l\pm4$, 
originating from the repulsive term of the potential. Numerical values of $a$ and the relative dipolar strength $\epsilon_{\text{dd}}=r_*/a$, 
where $r_*= m d_{\text{eff}}^2/\left(4 \pi \epsilon_0 \hbar^2\right)$ is the effective characteristic dipole-dipole distance, and
$d_{\text{eff}}= d/\sqrt{12(1+\delta_r^2)}$ is the effective dipole moment, for different parameters of the molecular interaction have been given in Ref.~\cite{Lang}.

\section{Equilibrium properties: Bogoliubov approach}  \label {Bogo}

In the momentum space, the Hamiltonian (\ref {ham}) can be written in terms of the creation, $\hat a^\dagger_{\bf k}$, and annihilation, $\hat a_{\bf k}$, operators 
by expanding the field operators $\hat\psi(\mathbf{r})=  (1/\sqrt{{\cal V}}) \sum_{\mathbf k} e^{i \mathbf k \cdot \mathbf r} \hat a_{\bf k}$, and 
$\hat\psi^\dagger(\mathbf{r})= (1/\sqrt{{\cal V}}) \sum_{\mathbf k} e^{-i \mathbf k \cdot \mathbf r} \hat a^\dagger_{\bf k}$, where ${\cal V}$ is a quantization volume. This gives:
\begin{align}\label{ham1}
&\hat H\!\!=\!\!\sum_{\bf k}\! E_k\hat a^\dagger_{\bf k}\hat a_{\bf k}\! 
+\!\frac{1}{2\cal V}\!\!\sum_{\bf k,\bf q,\bf p}\!\!
V^{\text{eq}}({\bf p})\hat a^\dagger_{\bf k\!+\!\bf q} \hat a^\dagger_{\bf k\!-\!\bf q}\hat a_{\bf k\!+\!\bf p}\hat a_{\bf k\!-\!\bf p},
\end{align}
where the superscript "eq" denotes the equilibrium state, $E_k=\hbar^2k^2/2m$ is the energy of a free particle,  and the interaction potential in momentum space is given by 
\begin{equation}\label{scam}
 V^{\text{eq}}(\mathbf k)= \int d^3 r e^{ - i \mathbf k \cdot \mathbf r}  V(\mathbf r).
\end{equation}
This integral can be evaluated employing the partial wave expansion for the plane wave $e^{ - i \mathbf k \cdot \mathbf r}= 4\pi \sum_{l,m} i^l j_l (kr)Y_l^{m*}(\theta,\phi) Y_l^m (\theta,\phi) $,
where $j_l (x)$ are spherical Bessel functions,
the orthonormality of the spherical harmonics $\int  d\theta d\phi  \sin \theta \,Y_l^{m*}(\theta,\phi)  Y_{l'}^{m'} (\theta,\phi)= \delta _{ll'} \delta_{mm\rq{}}$,
and the identity $\int_0^{\infty} x^2 J_l(x) / x^s = 2^{2 - s} \Gamma[ (3 + l - s)/2]/\Gamma[(-1 + l+ s)/2]$ with $s>3/2$.
However,  for $l=l'=1$, $V_{11,m}$ has a divergent term arising from $1/r^6$ shielding potential. To overcome such a divergence we follow the method outlined in Ref.~\cite{Deng}
and introduce a short-range cutoff, $r_{\text{UV}}$, on the lower integration limit. As a result the divergent term is $\propto k^2/r_{\text{UV}}$.
The convergence of the numerical solution requires the limit $r_{\text{UV}} \rightarrow 0$ \cite{Deng}.

In the weakly interacting regime we may use the Bogoliubov approach which  assumes that the ground state contains most of the molecules.
Applying the Bogoliubov transformations \cite{Bog}: 
\begin{equation}\label {BogTrans} 
\hat a_{\bf k}= u_k^{\text{eq}} \hat b_{\bf k}-v_k^{\text{eq}} \hat b^\dagger_{-\bf k},
\end{equation}
where $\hat b^\dagger_{\bf k}$ and $\hat b_{\bf k}$ are operators of elementary excitations.
The Bogoliubov functions $ u_k^{\text{eq}},v_k^{\text{eq}}$ in the equilibrium state are expressed in a standard way:
\begin{equation}\label {Bogfunc}
 u_k^{\text{eq}},v_k^{\text{eq}}= \left(\sqrt{\varepsilon_k^{\text{eq}}/E_k}\pm\sqrt{E_k/\varepsilon_k^{\text{eq}}}\right)/2,
\end{equation}
where the equilibrium Bogoliubov excitation energy is given by 
\begin{equation}\label {Bspec}
\varepsilon_{k}^{\text{eq}}=\sqrt{E_k^2+ 2n V^{\text{eq}}({\bf k})E_k},
\end{equation}
where $n=N/{\cal V}$ is the total density.
The final bilinear Hamiltonian of a BEC of MSPMs reads
\begin{equation}\label{ham2}
\hat H = E+\sum\limits_{\mathbf k} \varepsilon_k^{\text{eq}}\hat b^\dagger_{\bf k}\hat b_{\bf k},
\end{equation}
where $E=\tilde V^{\text{eq}}(|\mathbf k|=0) n N /2+\sum_{\bf k} [\varepsilon_k^{\text{eq}} -E_k-n V^{\text{eq}}({\bf k})]/2$ is the ground-state energy.

The noncondensed and the anomalous densities are  defined as : $\tilde{n}={\cal V}^{-1}\sum_{\mathbf k} n_{\mathbf k}$, and $\tilde{m}=-{\cal V}^{-1}\sum_{\mathbf k} m_{\mathbf k}$, respectively, where $n_{\mathbf k}= \langle\hat a^\dagger_{\bf k}\hat a_{\bf k}\rangle$ and 
$m_{\mathbf k}= \langle\hat a_{\bf k}\hat a_{-\bf k}\rangle$, are the normal and the anomalous distributions. 
A straightforward calculation yields in the thermodynamic limit:
\begin{equation}\label {nor}
\tilde{n}=\frac{1}{2}\int \frac{d^3k} {(2\pi)^3} \frac{E_k+V^{\text{eq}}({\bf k}) n} {\varepsilon_k^{\text{eq}}},
\end{equation}
and
\begin{equation}\label {anom}
\tilde{m}=-\frac{1}{2}\int \frac{d^3k} {(2\pi)^3} \frac{V^{\text{eq}}({\bf k}) n} {\varepsilon_k^{\text{eq}}}.
\end{equation}
The anomalous density is an important quantity in molecular BECs since it describes the correlations between the condensed and noncondensed molecules.

The normal and the anomalous distribution manifest in the EoS as \cite{LHY,Boudj5} 
\begin{align} \label{chim0}
\mu&=\tilde V^{\text{eq}}(|\mathbf k|=0) n + \mu_{\text{LHY}}, \\
&=\tilde V^{\text{eq}}(|\mathbf k|=0) n +\int \frac{d^3k} {(2\pi)^3} \tilde V^{\text{eq}}(\mathbf k) \big(\tilde n_k +\tilde m_k \big). \nonumber
\end{align}
The last term is the Lee-Huang-Yang (LHY) quantum corrections to the chemical potential.

Another important quantity is the second-order (pair) correlation function which can also be defined through the normal and anomalous distributions as:
\begin{equation}\label {corr}
g_2(r)=n^2+n\int \frac{d^3k} {(2\pi)^3} \left(2n_{\mathbf k} +m_{\mathbf k}\right) e^{i \mathbf k \cdot \mathbf r},
\end{equation}
it enables us to describe the internal structure of molecular BECs.

In the absence of the shielding term  $C_6=0$, the quantum depletion, the anomalous density and the LHY-correction EoS turn out be  given as:
$\tilde{n}=  {\cal Q}_3(-\epsilon_{\text{dd}}^{\text{eq}})/[3\pi^2 ({ \xi}^{\text{eq}})^3]$, $\tilde{m}={\cal Q}_3(-\epsilon_{\text{dd}}^{\text{eq}})/[\pi^2 (\xi^{\text{eq}})^3]$,
and $\mu_{\text{LHY}}=4{\cal Q}_5(-\epsilon_{\text{dd}}^{\text{eq}})/[3\pi^2 (\xi^{\text{eq}})^3]$, 
where $\epsilon_{\text{dd}}^{\text{eq}}=r_*^{\text{eq}}/a^{\text{eq}}$, $a^{\text{eq}}$ is the equilibrium $s$-wave scattering length,
$r_*^{\text{eq}}= m (d_{\text{eff}}^{\text{eq}})^2/\left(4 \pi \epsilon_0 \hbar^2\right)$ is the equilibrium dipolar distance, with 
$d_{\text{eff}}^{\text{eq}}$ being the equilibrium effective dipole moment, $\xi^{\text{eq}}= 1/\sqrt{4\pi n a^{\text{eq}}}$ is the healing length, and  the functions 
${\cal Q}_j(-\epsilon_{\text{dd}}^{\text{eq}})=(1+\epsilon_{\text{dd}}^{\text{eq}})^{j/2} {}_2\!F_1\left(-\frac{j}{2},\frac{1}{2};\frac{3}{2};\frac{3\,\epsilon_{\text{dd}}^{\text{eq}}}{1+\epsilon_{\text{dd}}^{\text{eq}}}\right)$ with  $j=3, 5$ \cite{lime,Boudj2},  
represent the DDI contribution to the condensate depletion and to the EoS, respectively. Here  ${}_2\!F_1(\alpha,\beta; \gamma;x)$ is the hypergeometric function.
The functions ${\cal Q}_j(-\epsilon_{\text{dd}}^{\text{eq}})$ are real for $-1<\epsilon_{\text{dd}}^{\text{eq}} \leq 1/2$, and become imaginary elsewhere,
they reach their maximum for $\epsilon_{\text{dd}}^{\text{eq}} \approx -1$.
Compared to the conventional DDI correction,  the obtained results differ by a minus sign in the argument of the ${\cal Q}_j$,
which may substantially change the real and imaginary contributions of ${\cal Q}_j$.
Note that the ultraviolet divergence appearing in the anomalous density (\ref{anom}) has been circumvented 
by introducing the Beliaev-type second-order coupling constant \cite {Beliaev,lime,Boudj2}.
Similarly to the case of atomic dipolar and nondipolar BECs \cite{Yuk},  the anomalous density of molecular condensates is three times larger than the noncondensed density. 
Both quantities are much smaller than total density ($\tilde{m}=3\tilde{n} \ll n$). Therefore,  the condition for the application of the Bogoliubov theory is well satisfied.

In the presence of the full molecular potential, obtaining exact analytical solutions of the integrals (\ref{nor})-(\ref{corr}) 
is not trivial due to the divergence arising from the short-range shilding part of the potential.
To handle such a dilemma, we perform numerical integration based on the aforementioned regularization scheme \cite{Deng}.
A converged solution is achieved when $r_{\text{UV}}$ is much smaller than the size of the shielding core.
For the length and the energy scales of the system,  we choose the dipolar length, $r_*^{\text{eq}}$, and the energy $E_*^{\text{eq}}=\hbar^2 / [m (r_*^{\text{eq}})^2]$, respectively.
We introduce the dimensionless parameters $\tilde C_n^{\text{eq}}= C_n/ [E_*^{\text{eq}} (r_*^{\text{eq}})^n]$.
In our numerical simulation we set $\tilde C_3^{\text{eq}}=1$ and  vary $\tilde C_6^{\text{eq}}$.
Experimentally, the effective dipole moment, the $s$-wave scattering length, the dipolar relative strength and hence the shielding parameter, $\tilde C_6$ can be tuned  
by adjusting the relative detuning, $\delta_r= |\delta|/\Omega$ \cite{Lang}. 

To be concrete we will use in what follows the parameters for bosonic NaK molecules \cite{KK} with mass $m=62$ amu, rotational constant $B/\hbar=2\pi\times 2.089$ GHz, 
and a fixed Rabi frequency of $\Omega/\hbar = 2\pi\times 10$ MHz \cite{Schind}.
The typical size of the shielding core is around $10^3 a_0$ (with $a_0$ the Bohr radius).

\begin{figure}
\includegraphics[scale=0.45] {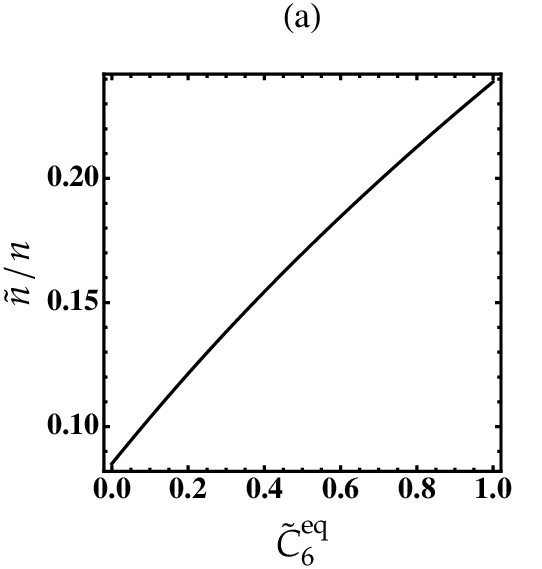}
\includegraphics[scale=0.45] {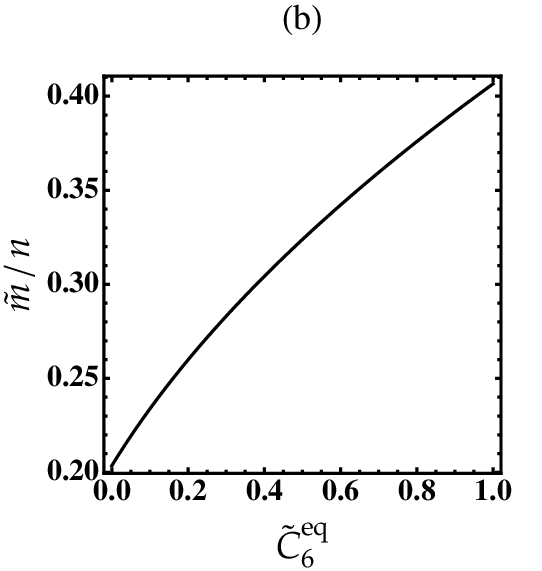}
\includegraphics[scale=0.45] {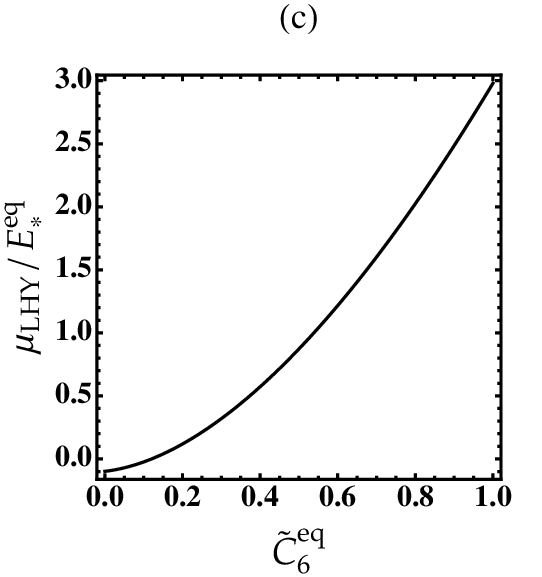}
\includegraphics[scale=0.46] {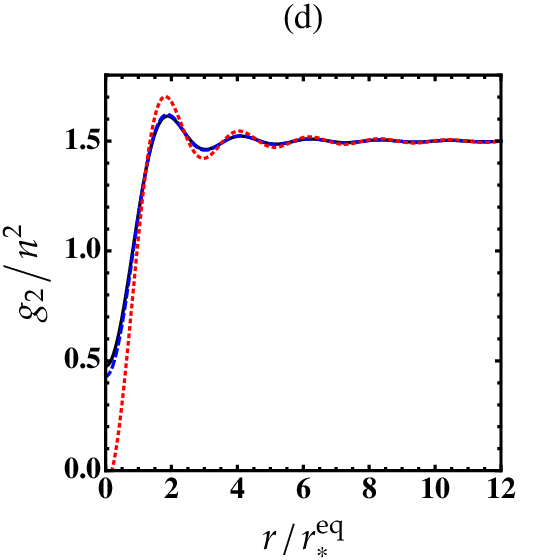}
 \caption{(a) Quantum depletion, $\tilde n/n$, from Eq.~(\ref{nor}) as a function of $\tilde C_6^{\text{eq}}$.
(b) Anomalous fraction, $\tilde m/n$, from Eq.~(\ref{anom}) as  a function of $\tilde C_6^{\text{eq}}$.
(c) LHY-corrected EoS from Eq.~(\ref{chim0})  as  a function  of $\tilde C_6^{\text{eq}}$.
(d) Second-order correlation function from Eq.~(\ref{corr}) for different values  of $\tilde C_6^{\text{eq}}$.
Black line: $\tilde C_6^{\text{eq}}=0.001$, blue-dashed line: $\tilde C_6^{\text{eq}}=0.01$, and red-dotted line: $\tilde C_6^{\text{eq}}=0.1$.}
 \label{EqCDs}
\end{figure}

We plot in Fig.~\ref{EqCDs}, the quantum depletion (\ref{nor}), the anomalous fraction (\ref{anom}), the LHY-corrected EoS (\ref{chim0}),
and the second-order correlation function (\ref{corr}) as a function of the dimensionless shielding term, $\tilde C_6^{\text{eq}}$.
As shown in Figs.~\ref{EqCDs} (a) and (b), both $\tilde n/n$ and $\tilde m/n$ increase monotonically with $\tilde C_6^{\text{eq}}$ revealing that 
a large number of molecules is spreaded out of the condensate when the shielding core term becomes important.
This, in turn, decreases the condensed fraction.
The LHY-corrected EoS is negative only for small $\tilde C_6^{\text{eq}}$ as seen in Fig.~\ref{EqCDs} (c) which could be a signature of the formation of a self-bound state.
In contrast, for a strong shielding potential, $\mu_{\text{LHY}}$ possesses positive values indicating the formation of a stable molecular gas.
Figure \ref{EqCDs} (d) depicts that the normalized second-order correlation function, $g_2(r)/n^2$, increases linearly at short distances, $r\simeq r_*^{\text{eq}}$ 
until it reaches its maximum, then it saturates for large distances, $r \gtrsim r_*^{\text{eq}}$ causing the condensed molecules to become less correlated.
As the shielding potential increases (i.e. the mean interparticle distance is reduced), $g_2(r)/n^2$ generates strong antibunching effects.
Furthermore, the existence of oscillations in the correlation function is attributed to the extra quasiparticle excitation induced by the shielding interactions.

\section{Quench dynamics} \label{QDyn}

We now use the TDBA for studying the dynamics induced by the interaction quench of weakly interacting molecular BECs.
The TDBA is a generalization of the standard Bogoliubov approximation, where the canonical transformation (\ref{BogTrans}) becomes time-dependent \cite{Natu}
\begin{equation}\label {trans}
 \hat a_{\bf k} (t)= u_k(t) \hat b_{\bf k}-v_k(t) \hat b^\dagger_{-\bf k}, 
\end{equation}
where $u_k(t)$ and $v_k (t)$ are time-dependent Bogoliubov amplitudes acquiring complex values and satisfying the standard normalization condition: $|u_k(t)|^2-|v_k(t)|^2=1$.
Their equations of motion are obtained from the Heisenberg equations of motion for $a_{\mathbf k}$ \cite{Natu}
\begin{align} \label{BdGE22}
i
		\begin{pmatrix}  
		\dot {u}_{k} (t)\\
		\dot {v}_{k}(t)\\
		\end{pmatrix}
\\
&=
&\begin{pmatrix}
		E_k+ V (\mathbf k) n & V (\mathbf k) n&\\
		-V (\mathbf k)  n &- E_k- V (\mathbf k) n)&\\
		\end{pmatrix}
		\begin{pmatrix}
		u_{k}(t)\\
		v_{k} (t)\\
		\end{pmatrix}.
		\nonumber
\end{align} 
Note that the time evolution of $u_{k}$ and $v_{k}$ depends only on the final Hamiltonian parameters since  we deal with a sudden quench.
The integration of the linear Eqs.~(\ref{BdGE22}) gives:
\begin{align} \label{BdGE2}
		\begin{pmatrix} 
		u_{k} (t)\\
		v_{k}(t)\\
		\end{pmatrix}
&= \bigg[\cos (\varepsilon_k t) \mathbb{1} -i \frac{\sin (\varepsilon_k t) } {\varepsilon_k^2} \\
&\times\begin{pmatrix}
		E_k+ V (\mathbf k) n & V (\mathbf k) n&\\
		-V (\mathbf k)  n &- E_k- V (\mathbf k) n)&\\
		\end{pmatrix}
		\bigg]
		\begin{pmatrix}
		u_{k}^{\text{eq}}\\
		v_{k}^{\text{eq}}\\
		\end{pmatrix} ,
		\nonumber
\end{align} 
where $\mathbb{1}$ is the identity matrix, $ u_k^{\text{eq}},v_k^{\text{eq}}$ are defined in (\ref{Bogfunc}) and 
\begin{equation}\label {TDspec}
\varepsilon_k=\sqrt{E_k^2+ 2 nV(\mathbf k) E_k},
\end{equation}
is the final Bogoliubov excitations spectrum.  

At zero temperature, the time-dependent noncondensed and anomalous densities are then given by $\tilde{n} (t)= \sum_k \tilde n_k(t) =\sum_k |v_k(t)|^2$ and 
$\tilde{m}(t)=\sum_k \tilde m_k (t)=-\sum_k \left[u_k (t) v_k^*(t)+u_k^*(t) v_k(t)\right]$, respectively. Using  the time-dependent solutions (\ref{BdGE2}), we get:
\begin{equation}\label {TDnor}
\tilde{n} (t)=  \tilde{n}+ \int \frac{d^3k}{(2\pi)^3} V(\mathbf k)[V(\mathbf k)- V^{\text{eq}}(\mathbf k)] n^2 E_k \frac{\sin^2 (\varepsilon_k t)}{\varepsilon_k^2 \varepsilon_k^{\text{eq}}},
\end{equation}
and
\begin{align}\label {TDanom}
\tilde{m}(t)&=\tilde{m} - 2\int \frac{d^3k}{(2\pi)^3}\frac{\sin^2 (\varepsilon_k t)}{\varepsilon_k^2\varepsilon_k^{\text{eq}}}\\
&\times\bigg\{ E_k^2 V(\mathbf k) n +E_k V(\mathbf k) n^2 [ V(\mathbf k)+V^{\text{eq}} (\mathbf k)] \bigg\} \nonumber.
\end{align}
These expressions show that the density of noncondensed molecules and anomalous correlations 
increase (decrease) depending on whether $V^{\text{eq}}> V$ or $V^{\text{eq}}<V$. For $V=0$,  there are no excitations created or destroyed in the system.

Equations (\ref{TDnor}) and (\ref {TDanom}) must satisfy the equality \cite{Boudj2,Boudj4}
\begin{equation}\label {TDheis}
I_k(t)=[2\tilde{n}_k(t)+1]^2-|2\tilde{m}_k(t)|^2,
\end{equation}
where $I$ accounts for the variance of the number of noncondensed particles (i.e. the condensate fluctuation).
Equation (\ref {TDheis}) clearly shows that $\tilde{m}$ is larger than $\tilde{n}$ at low temperature.

The equal time second-order correlation function affords another interesting description of the quenched dynamics of ultracold shielded molecules.
Its dynamics is given by: 
$g_2(r,t)=n^2+n\int \frac{d^3k} {(2\pi)^3} \left(2n_{\mathbf k} (t) +m_{\mathbf k} (t)\right) e^{i \mathbf k \cdot \mathbf r}$. 
Upon introducing the time-dependent Bogoliubov amplitudes (\ref{BdGE2}), one finds
\begin{align}\label {TDcorr}
g_2(r,t)&=g_2(r) - 2n\int \frac{d^3k} {(2\pi)^3} \frac{E_k \sin^2 (\varepsilon_k t)}{\varepsilon_k^2 \varepsilon_k^{\text{eq}}}  \\
&\times V(\mathbf k) n [ E_k+nV^{\text{eq}} (\mathbf k)] e^{i \mathbf k \cdot \mathbf r}. \nonumber
\end{align}
The minus sign on the right-hand side of Eq.~(\ref{TDcorr}) which comes from the contribution of the anomalous term, $\tilde m_k$, indicates that in the case of a molecular BEC,
the probability of finding a particle at time $t$ and in a distance $r$ becomes smaller than that of an ideal system.

Evidently, a variation in the microwave-dressed molecular interactions may significantly modify the excitation spectrum, marking deviations in the quantum depletion, 
the condensate fluctuations, and the pair correlations.
In what follows we will model the quench of the microwave-dressed molecular interactions using two configurations.

\subsection{Quench from $V^{\text{eq}}=0$ to $V (\mathbf k)>0$} \label{QA}

Let us start with the popular case and suppose a quench from $V^{\text{eq}}(\mathbf k)=0$ when $t\leq 0$ to $V (\mathbf k)>0$ at time $t >0$.
In such a case, $\tilde n=\tilde m=0$, hence the time-dependent noncondensed and anomalous densities reduce to 
\begin{equation}\label {TDnor1}
\tilde{n} (t)=  \int \frac{d^3k}{(2\pi)^3} n^2 V^2(\mathbf k) \frac{\sin^2 (\varepsilon_k t)} {\varepsilon_k^2},
\end{equation}
\begin{align}\label {TDanom1}
\tilde{m}(t)=-2\int \frac{d^3k}{(2\pi)^3}\frac{\sin^2 (\varepsilon_k t)}{\varepsilon_k^2}\bigg\{ V(\mathbf k)n [ E_k+ V(\mathbf k) n] \bigg\},
\end{align}
One can compute the condensate fluctuation and the correlation function by directly inserting the normal and anomalous distributions into Eqs.~ (\ref{TDheis}) and (\ref{TDcorr}).
This gives:
\begin{align}\label {TDheis1}
I(t)&= \int \frac{d^3k}{(2\pi)^3} n^2 V^2(\mathbf k) \frac{\sin^2 (\varepsilon_k t)}  {\varepsilon_k^4}\\
&\times[ \varepsilon_k^2- (4\varepsilon_k^2+3 n V(\mathbf k)) \sin^2 (\varepsilon_k t)], \nonumber
\end{align}
and 
\begin{align}\label {TDcorr1}
g_2(r,t)=n^2 - 2\int \frac{d^3k} {(2\pi)^3} n^2 V(\mathbf k)\frac{E_k \sin^2 (\varepsilon_k t)}{\varepsilon_k^2}  e^{i \mathbf k \cdot \mathbf r}. 
\end{align}
From now on, we introduce the characteristic relaxation time following the quench, $\tau =\hbar/E_*$,  the energy, $E_*=\hbar^2 / (m r_*)$, 
and the characteristic dipolar length, $r_*$, as  the units for time, energy, and length, respectively.

At long times, the main contribution to the above integrals comes from the low momentum limit where the Bogoliubov dispersion relation becomes linear: $\varepsilon_k= \hbar c_s(\theta) k$,
where $c_s(\theta)= \sqrt{ n V (|\mathbf k |=0)/m}$  is the sound velocity. Within this, we obtain:
\begin{equation}\label {TDnor2}
\tilde{m} (t \rightarrow \infty) \sim 2 \tilde{n}(t \rightarrow \infty)=  \frac{1}{8\pi \xi^3} f(t),
\end{equation}
where $\xi= 1/\sqrt{4\pi n a}$, and 
\begin{align}
f(t)= \int_0^{\pi}  \sin \theta  d \theta  \bigg\{\frac{1- e^{-4 t/\left[\tau \left(1+ \epsilon_{\text{dd}} (1-3\cos^2\theta)\right) \right]} } 
{ \left[1+ \epsilon_{\text{dd}} (1-3\cos^2\theta)  \right]^{3/2}} \bigg\}.
\end{align}
On timescales $t\simeq \tau \left(1+ \epsilon_{\text{dd}} (1-3\cos^2\theta)\right)$, one gets 
$\tilde{m} (t \rightarrow \infty) \sim 2 \tilde{n}(t \rightarrow \infty)= {\cal Q}_3 (- \epsilon_{\text{dd}})/ (4\pi \xi^3)$ with  $-1<\epsilon_{\text{dd}} \leq 1/2$.
Remarkably, $\tilde{m} (t \rightarrow \infty)$ and $\tilde{n}(t \rightarrow \infty)$ for MSPM BECs are larger than the equilibrium values
of 3D atomic BECs with DDI, $\tilde n= 3\tilde m= {\cal Q}_3 (\epsilon_{\text{dd}})/(3\pi^2\xi^3)$ \cite{Boudj2}.
In the absence of the microwave-dressed molecular interactions, we find $\tilde{m} (t) \sim 3 \tilde{n}(t)= ( 1- e^{-4t/\tau})/\big(4\pi \xi^3\big)$, 
showing that both the noncondensed and anomalous densities grow linearly at short times, while they saturate at $t \sim \tau $.
In the long time limit, both quantities are larger than their equilibrium values for nondipolar Bose gases, $\tilde n= 3\tilde m= 1/(3\pi^2\xi^3)$ \cite {Boudj2},
which means that the creation of pair excitations is important in such a regime. 

In the long-time regime, the condensate fluctuation asymptotes to:
\begin{equation}\label {TDheis1}
I (t \rightarrow \infty)-1 \sim -\frac{3}{16\pi \xi^3} g(t),
\end{equation}
where 
\begin{align}
&g(t)=  \int_0^{\pi} \bigg \{5 + \frac{8t}{\tau \left(1+ \epsilon_{\text{dd}} (1-3\cos^2\theta)\right)} \\
&+ \bigg[-5+\frac{4t}{\tau\left(1+ \epsilon_{\text{dd}} (1-3\cos^2\theta)\right)} \bigg] e^{-4t/\left[\tau \left(1+ \epsilon_{\text{dd}} (1-3\cos^2\theta)\right) \right]} \bigg\}  \nonumber\\
& \times\frac{\sin \theta  d \theta  } { \left[1+ \epsilon_{\text{dd}} (1-3\cos^2\theta)  \right]^{3/2}} \nonumber.
\end{align}
The additional terms appearing in the function $g(t)$ comes from the coupling of the normal and anomalous correlations. 
For $t\simeq \tau \left(1+ \epsilon_{\text{dd}} (1-3\cos^2\theta)\right)$, the condensate fluctuation takes the form:
$I (t \rightarrow \infty)-1  \sim -(39/8\pi \xi^3)  {\cal Q}_3 (- \epsilon_{\text{dd}})$.
In the case of a BEC with a pure contact interactions, one has
\begin{equation}\label {TDheis2}
I (t \rightarrow \infty)-1 \sim \frac{3}{8\pi \xi^3}  \bigg [5 + \frac{8t}{\tau } + \left(-5+\frac{4t}{\tau} \right) e^{-4t/\tau } \bigg]. 
\end{equation}
We see that the condensate fluctuation decays exponentially for both dipolar and nondipolar atomic BECs at long times and does not recover its equilibrium value.

\begin{figure}
\includegraphics[scale=0.44] {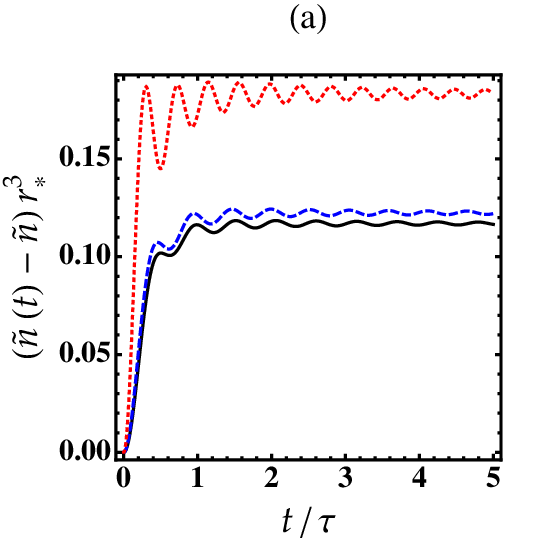}
\includegraphics[scale=0.44] {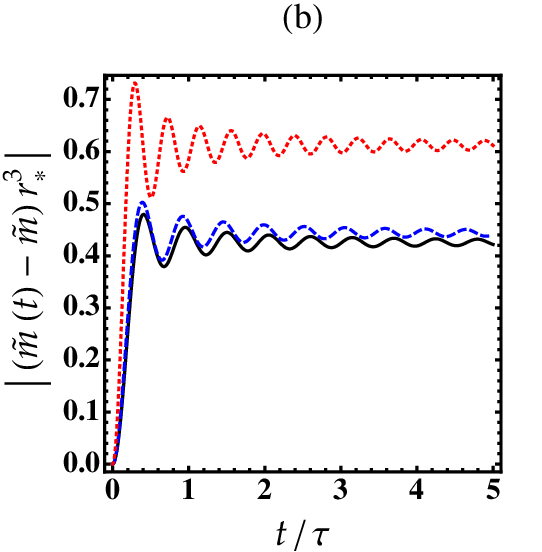}
\includegraphics[scale=0.44] {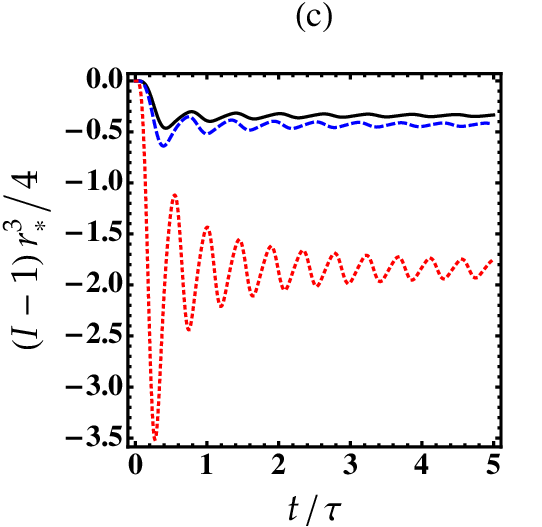}
\includegraphics[scale=0.44] {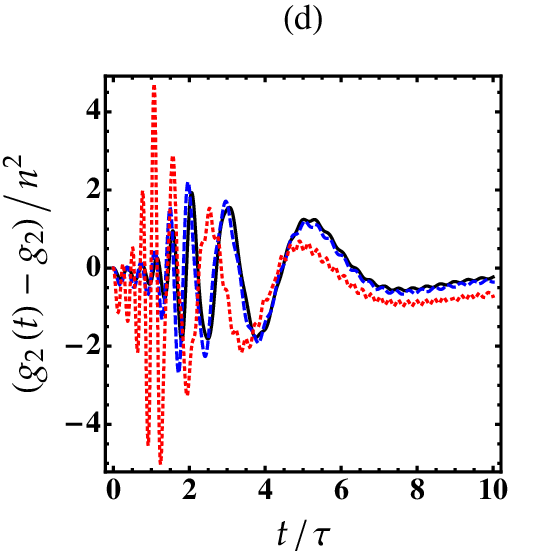}
 \caption{ (a) Time-evolution of the noncondensed density normalized to the equilibrium depletion, $\tilde n (t)-\tilde n$ for different values of $\tilde C_6$.
(b) Time-evolution of the absolute value of the anomalous density  normalized to the equilibrium value, $|\tilde m (t)-\tilde m|$, for different values of $\tilde C_6$.
(c) Time-evolution of the condensate fluctuation, $(I(t)-1)/4$, for different values of $\tilde C_6$.
(d) Normalized second-order correlation function, $(g_2(r, t) - g_2)/n^2$,  for different values of $\tilde C_6$ with $r/r_*=12$.
Black line: $\tilde C_6=0.001$, blue-dashed line: $\tilde C_6=0.01$, and red-dotted line: $\tilde C_6=0.1$.}
 \label{CDs}
\end{figure}

Next, we consider the dynamics of the full time-dependent Eqs.~(\ref{TDnor1})-(\ref{TDcorr1}) by implementing a time-propagation scheme of 
the equilibrium solutions with the aforementioned truncation.
In Fig.~\ref{CDs} we plot the time-dependent depletion,  anomalous density, condensate fluctuation, and correlation function  for different values
of the dimensionless shielding term, $\tilde C_6= E_* r_* C_6$.

Figures \ref{CDs}. (a) and (b) show that at short times  $t \lesssim 0.5 \,\tau$ both $\tilde n (t)$ and $\tilde m (t)$ increase sharply  then they saturate at long times, $t > 0.5 \,\tau$ 
regardless of the interaction strength, $\tilde C_6$. The nonoscillatory behavior of this growth originates most probably from the very low momenta.
For weak shielding interactions, $\tilde C_6< 0.01$, $\tilde n (t)$ and $\tilde m (t)$ develop damped oscillations.
With an increasing shielding interaction, the number of new excitations created is large, causing pronounced and undamped oscillations. 
Remarkably, the time-dependent anomalous density remains larger than the noncondensed density at any time and for any value of the shielding term.
The same behavior holds for equilibrium dipolar and nondipolar atomic BECs  \cite {Boudj2,Boudj5}.

Figure \ref{CDs}. (c) depicts that the normalized condensate fluctuation, $(I(t)-1)/4$,  decreases with $\tilde C_6$ and exhibits damped oscillations at long times. 

In Fig.~\ref{CDs} (d) one can clearly identify two phases of evolution of the normalized second-order correlation function.
In the first phase at times $t \lesssim 2\tau$,  $(g_2(r, t) - g_2)/n^2$ experiences fast oscillations
which can be attributed to the spreading of rapid quasiparticles that acquire a high kinetic energy after the quench.
For relatively strong shielding interactions, $\tilde C_6 \simeq 0.1$,  $(g_2(r, t) - g_2)/n^2$ exhibits high amplitude oscillations and can take negative values (i.e. anti-correlation).
Similar behavior has been observed for the quench dynamics of a Rydberg-dressed BEC \cite{Cormack}.
In the long time limit, $t > 2\tau$, the correlation function oscillates slowly with large width and tends to zero regardless of the value of $\tilde C_6 $. 
The slow oscillations  correspond most probably to the excitation of phonon modes.

\begin{figure}
\includegraphics[scale=0.8] {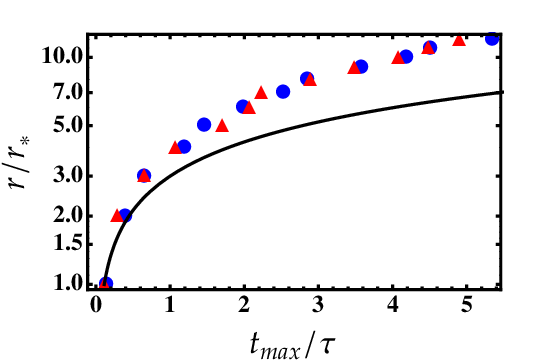}
 \caption{ Temporal location of the last maximum in the correlation function ($t_{\text{max}}$) for different values of $\tilde C_6$.
Triangles: $\tilde C_6=0.01$. Circles: $\tilde C_6=0.1$. Solid line shows  purely diffusive fit to the data.
Parameters are the same as in Fig.~\ref{CDs}.}
 \label{Ball}
\end{figure}

On longer time scales, $(g_2(r, t) - g_2)/n^2$ develops local maxima. The temporal location of the last maximum in the pair correlation function is shown in Fig.~\ref{Ball}.
We see that at short times, the short-range correlations dynamics is diffusive ($ r/r_* \simeq D (t/\tau)^{1/2}$) with a diffusion constant $D \approx 3.1$, 
while a ballistic motion is observed in the limit of long distances ($r>r_*$) and long times. This behavior holds for any value of the shielding term, $\tilde C_6$.
Note that such a crossover from the diffusive to the ballistic regime has been also found in the dynamics of a dilute Bose gas with a pure contact interaction \cite{Natu}.

\subsection{Quench from $V^{\text{eq}}>0$ to $V>0$} \label{QB}

\begin{figure}
\includegraphics[scale=0.46] {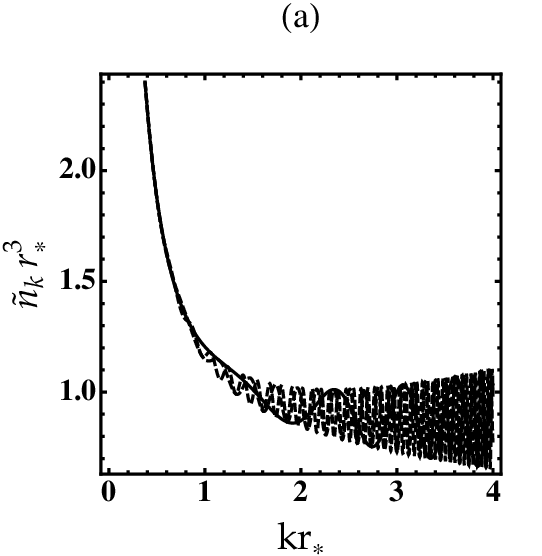}
\includegraphics[scale=0.46] {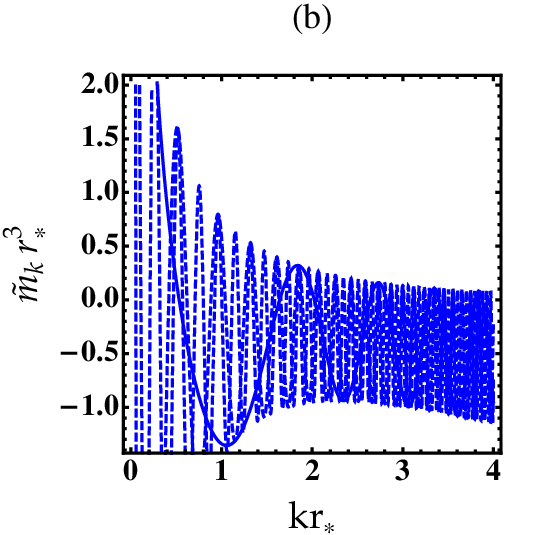}
 \caption{(a) Normal, $n_k$ and (b) anomalous, $m_k$ momentum distributions at different values of $t/\tau$ with $\tilde C_6^{\text{eq}}=0.1$ and $\tilde C_6=0.1$.
Solid lines: $t/\tau=1$. Dashed lines: $t/\tau=5$. Dotted lines: $t/\tau=10$.}
 \label{dist}
\end{figure}

Now we consider a quench described by the sudden change from some initial microwave-dressed molecular $V^{\text{eq}}>0$ to a final value $V>0$.
During the simulation process, the equilibrium dipolar and shielding parameters are fixed as: $\tilde C_3^{\text{eq}}=\tilde C_3=1$, and $\tilde C_6^{\text{eq}}=0.001$.

In Fig.~ \ref{dist} we plot the normal and anomalous momentum distributions at different times.
We see that they are strongly affected by the presence of the effective intermolecular potential. 
Both $\tilde n_k$ and $\tilde m_k$ oscillate rapidly in the high momenta regime and at long times.
A careful observation reveals that $\tilde m_k$ oscillates stronger than $\tilde n_k$ even  at low momenta (i.e. the phonon regime) 
giving rise to a nontrivial dynamical evolution of the noncondensed and anomalous densities.
Experimentally, the molecule distribution can be determined using the matter-wave refocusing technique \cite{Mur,Muk}.

\begin{figure}
\includegraphics[scale=0.45] {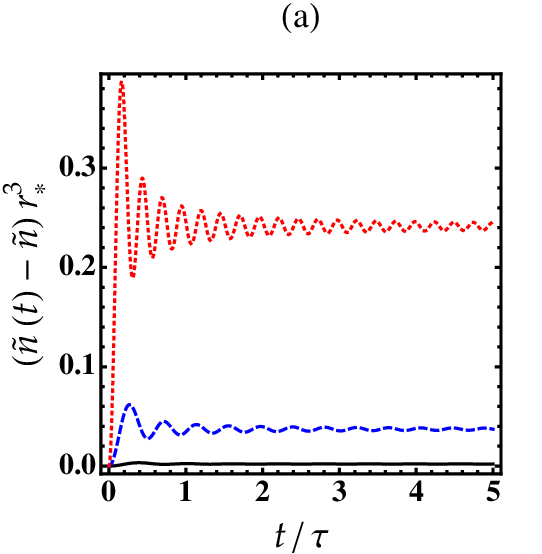}
\includegraphics[scale=0.44] {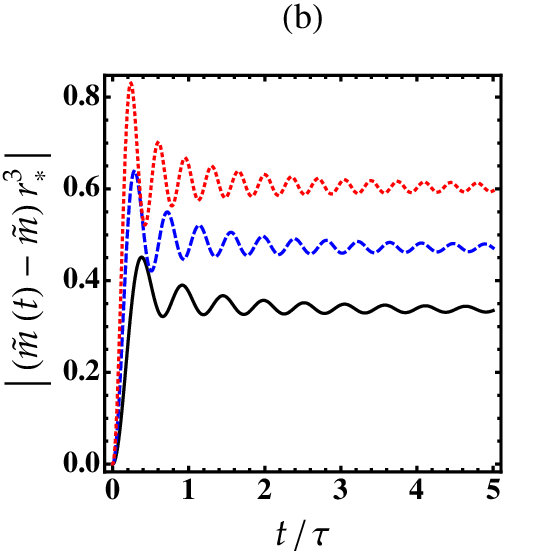}
\includegraphics[scale=0.45] {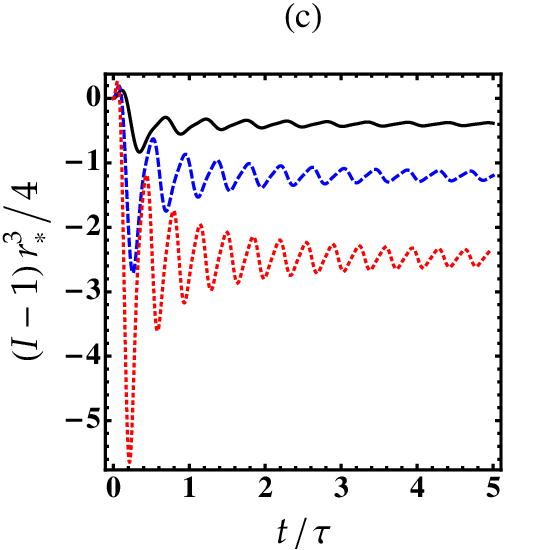}
\includegraphics[scale=0.44] {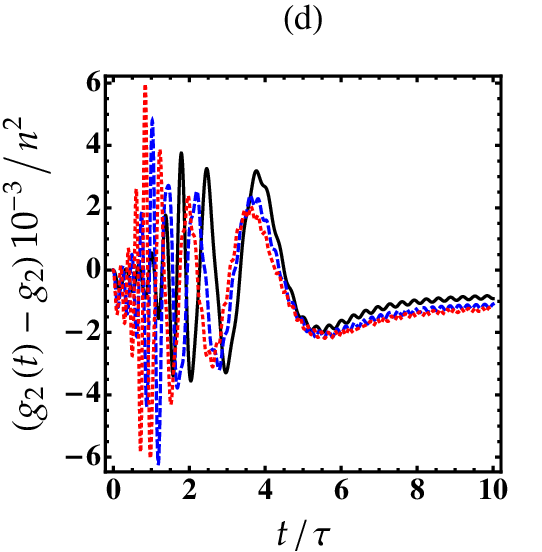}
 \caption{ (a) Time-evolution of the noncondensed density normalized to the equilibrium depletion, $\tilde n (t)-\tilde n$ for different values of $\tilde C_6$.
(b) Time-evolution of the absolute value of the anomalous density  normalized to the equilibrium value, $|\tilde m (t)-\tilde m|$, for different values of $\tilde C_6$.
(c) Time-evolution of the condensate fluctuation, $(I(t)-1)/4$, for different values of $\tilde C_6$.
(d) Normalized second-order correlation function, $(g_2(r, t) - g_2)/n^2$,  for different values of $\tilde C_6$ with $r/r_*=12$.
Black line: $\tilde C_6=0.01$, blue-dashed line: $\tilde C_6=0.1$, and red-dotted line: $\tilde C_6=0.2$.}
 \label{2CDs}
\end{figure}

The numerical results for the condensate depletion, the anomalous density, the condensate fluctuation and the second-order correlation function are shown in Fig.~\ref{2CDs}.
We observe that the quantum depletion and the anomalous density saturate towards $\sim 60\% $ and  $\sim 80\% $, respectively which are 
larger than their equilibrum states (see Figs.~\ref{2CDs} (a) and (b)).
Both quantities develop very damped oscillations compared to the previous case due to the competition of dipolar and shielding terms.
Therefore,  if the microwave shielding remains on, the molecule number decays owing to the dephasing of the microwave-dressed states.
After such a dephasing, the cloud becomes a mixture of molecules in the ground and excited rotational states.
The condensate fluctuation, $(I(t)-1)/4$, and the normalized second-order correlation function, $(g_2(r, t) -g_2)/ n^2$ exhibit almost the same behavior 
as in the previous case described in Sec.~\ref{QA} with a slight difference in the oscillation amplitude (see Figs.~\ref{2CDs} (c) and (d)). 
Consequently, the long-range correlations ($r>r_*$) can expand ballistically in the long-time regime regardless of the value of $\tilde C_6$.

\section{Self-consistent time-dependent Hartree-Fock-Bogliubov approximation}\label{HFB}

In this section we use the self-consistent TDHFB approach to describe the nonequilibrium properties of molecular BECs 
and compare the results with the TDBA findings.
The HFB theory is valid for both weak and strong interactions, at all times and at any temperatures  \cite{Boudj5,Boudj6, Yuk, Boudj7}, in contrast to
the standard TDBA which is valid provided that $\tilde n \ll n$ and $\tilde m \ll n$ (i.e. only for weakly interacting systems) and at short times.

\begin{figure}
\includegraphics[scale=0.46] {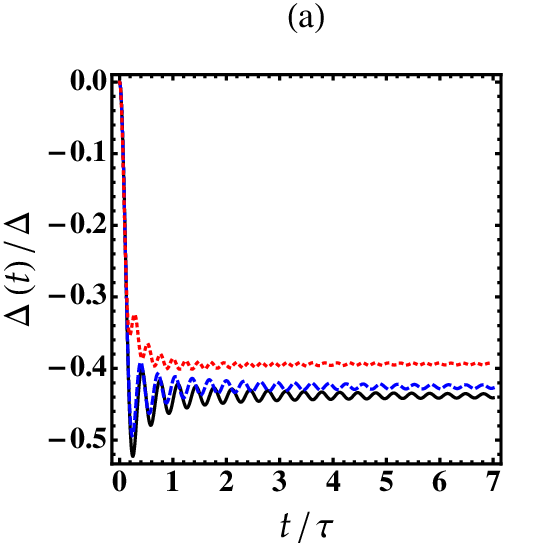}
\includegraphics[scale=0.45] {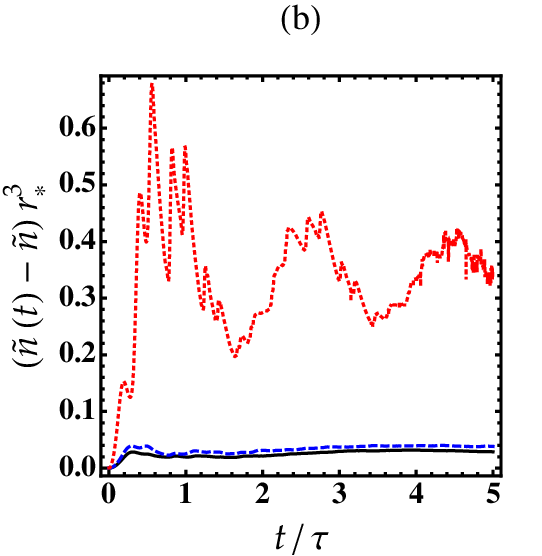}
\includegraphics[scale=0.45] {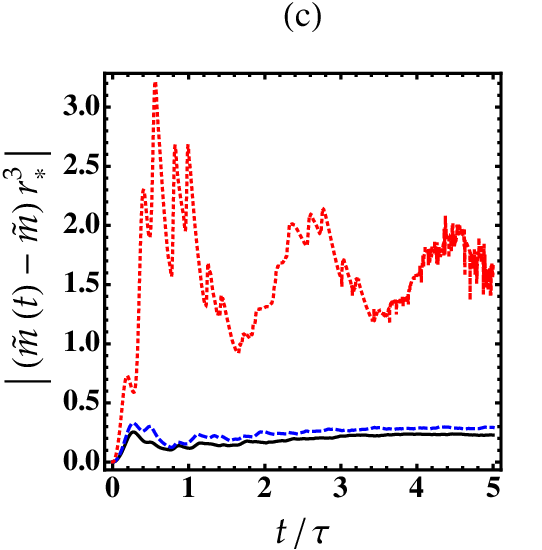}
\includegraphics[scale=0.45] {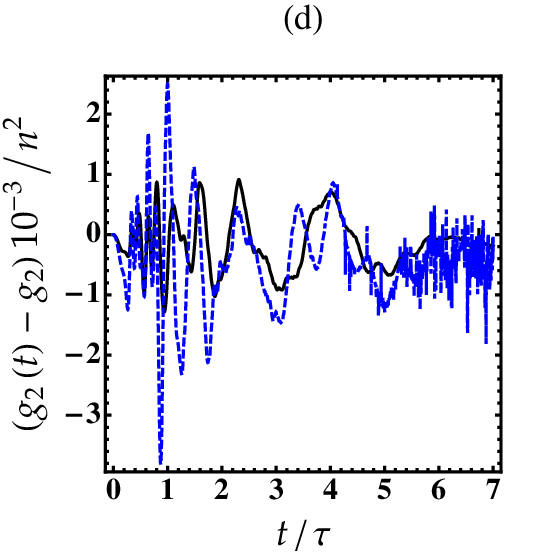}
 \caption{(a) Time-evolution of the normalized gap, $\Delta (t)/\Delta$, for different values of $\tilde C_6$.
(b) Time-evolution of the noncondensed density normalized to the equilibrium depletion, $\tilde n (t)-\tilde n$ for different values of $\tilde C_6$.
(c) Time-evolution of absolute value of the anomalous density  normalized to the equilibrium value, $|\tilde m (t)-\tilde m|$, for different values of $\tilde C_6$.
(d) Normalized second-order correlation function, $(g_2(r, t) - g_2)/n^2$,  for different values of $\tilde C_6$ with $r/r_*=12$.
Black line: $\tilde C_6=0.001$, blue-dashed line: $\tilde C_6=0.01$, and red-dotted line: $\tilde C_6=0.1$.}
 \label{hfbCDs}
\end{figure}

The full TDHFB theory requires that higher-order terms must be taken into account. 
Therefore, the Hamiltonian (\ref{ham1}) can be represented as a sum of five terms, $\hat H=\sum\limits_{j=0}^5\hat H^{(j)}$, depending on the powers of the operators,
$\hat a^\dagger_{{\bf k} \neq0}$, and $\hat a_{{\bf k} \neq0} $.
Following the same fashion, we obtain for the Bogoliubov quasiparticle amplitudes
\begin{equation}   \label{Bogfunc11}
u_k^2 = \frac{\omega_k+\varepsilon_k}{2\varepsilon_k}, \qquad
v_k^2 = \frac{\omega_k-\varepsilon_k}{2\varepsilon_k},
\end{equation}
and for the Bogoliubov excitations energy \cite{Boudj5,Yuk} 
\begin{equation} \label{Bog11}
\varepsilon_k =\sqrt{\omega_k^2 -\Delta_k^2},
\end{equation}
where 
$$
\omega_k \equiv E_k + n  \tilde V (|\mathbf k|=0) + n_c  \tilde V ({\bf k}) +\frac{1}{V} \sum_{p\neq 0} \tilde n_p  \tilde V ({\bf k}+{\bf p}) - \mu_1,
$$
and
$$
\Delta_k \equiv n_c  \tilde V ({\bf k}) +\frac{1}{V} \sum_{p\neq 0} \tilde m_p  \tilde V ({\bf k}+{\bf p}),
$$
with $n_c$ being the condensed density.\\
Evidently, the HFB spectrum (\ref{Bog11})  has a  gap in the limit of long wavelengths due to the inclusion of the anomalous correlations.
To solve this problem, we introduce the chemical potential $\mu_1$ as \cite {Boudj5,Yuk}:
\begin{equation}
\label{Chim1}
\mu_1= n  \tilde V(|\mathbf k|=0)+ \frac{1}{V}  \sum_{k\neq0} \tilde V ({\bf k}) (\tilde n_k -\tilde m_k).
\end{equation}
It is obvious that the chemical potential $\mu_1$ of Eq.(\ref{Chim1}) renders the spectrum (\ref{Bog11}) gapless satisfying the Hugenholtz-Pines theorem \cite{HP}. 
Importantly,  the gap, $\Delta_k$, which relies on the anomalous correlations and on the Bogoliubov excitations spectrum, becomes time dependent.
So, the solution of the above self-consistent equations requires the use of an iterative scheme.

For simplicity we consider  here a quench from $V^{\text{eq}}(\mathbf k)=0$ when $t\leq 0$ to $V (\mathbf k)>0$ at time $t >0$ and ignore some unimportant terms.
We then solve iteratively Eqs.~(\ref{TDnor}), (\ref{TDanom}), (\ref{TDcorr}), (\ref{Bogfunc11}), and (\ref{Bog11}) up to second-order in $\tilde n$ and $\tilde m$ 
following the method outlined in our recent work \cite{Boudj8}.  
The numerical results are shown in Fig.~\ref{hfbCDs}. \\
Figure \ref{hfbCDs} (a) shows that the normalized gap, $\Delta (t)/\Delta$, sharply decreases at short times and then it saturates at long times.
For relatively strong shielding interaction ($\tilde C_6=0.1$), the gap oscillates rapidly with negligible damping at long times.
To the best of our knowledge, the dynamics of the gap has never been analyzed before in the literature.
Remarkably,  the normal and anomalous densities become crucial compared to those of the TDBA due to the multiple counting of the interaction between 
condensed and noncondensed molecules arising from the pairing term 
which induces large-amplitude oscillations notably for large shielding interaction as seen in  Figs.~\ref{hfbCDs} (b) and (c). 
Unlike the TDBA results, the pair correlation function develops fast oscillations at both short and long times and then slow oscillations at intermediate times as demonstrated in Fig.~\ref{hfbCDs} (d).
Such oscillations become very rapid as the shielding interaction is increased and hence the quantum fluctuations become strong enough to destroy the ordering
signaling that the system undergoes a phase transition induced by quenching of long-range.

\section{Conclusions and outlook}\label{conc}
              
In this paper, we presented a comprehensive study of the nonequilibrium dynamics of homogeneous ultracold Bose gases of MSPMs
shedding light on how an interaction quench can modify the evolution of such systems.

In the equilibrium case, we showed that the quantum depletion and the anomalous density are significantly increasing with the shielding interaction leading to a
reduction of the condensed fraction. The EoS changes its character from negative to positive with the shielding interaction strength, 
and the pair correlation function presents an oscillatory behavior at long times.

On the other hand, an interaction quench is applied by switching on the interaction instantaneously, starting first  from
a noninteracting Bose gas and second from some initial value of intermolecular potential.
We showed that the depletion, the anomalous density, the condensate fluctuations and the pair correlation functions develop fast, slow, 
damped, and undamped oscillations depending on the strength of the shielding interaction.
Our results revealed also that the pair correlations expand diffusively at short times, crossing over to ballistic motion at long times.
In the case of a sudden quench from noninteracting Bose gas to a molecular condensate, we demonstrated analytically  that at long time scales the 
system supports a steady state that differs from its corresponding equilibrium value.
Moreover,  we found that the depletion, the anomalous density and the correlations computed by the self-consistent HFB theory
are larger than those of the TDBA and display persistent oscillations for relatively strong shielding interactions owing to the higher-order quantum fluctuations. 
The opposite situation holds for the gap parameter.

The present study presents an opportunity for exploring and probing the peculiar dynamics of correlations in uniform ultracold molecular gases of MSPMs 
following a sudden ramp of the system interaction. Experimentally, this can be readily achieved using high resolution imaging techniques \cite{Bakr,Sher}. 
Furthermore, the creation of new excitations due to interaction quench may lead to the simultaneous access to thermal molecular gases allowing us to 
measure the quantum depletion.  
In the case of atomic BECs, this can be realized employing the coherent two-photon Bragg scattering \cite{Lopes}.
When the interaction potential between two MSPMs is quenched,  the depletion extends over a large momentum range featuring strong correlations between opposite momenta 
(i.e. the anomalous density), enabling the detection of such anomalous correlations without the need
of a single-atom resolution used in atomic BECs \cite{Tenart}.

\section*{Acknowledgments}
We are grateful to Immanuel Bloch for valuable discussions.

\end{document}